\documentstyle[aps,epsfig]{revtex}
\begin{document}
\twocolumn[\hsize\textwidth\columnwidth\hsize\csname @twocolumnfalse\endcsname

\draft
 
\title{
Dynamic correlations in doped 1D Kondo insulator\\ - Finite-$T$ DMRG study -
}
\author{
Naokazu {\sc Shibata} and Hirokazu {\sc Tsunetsugu}
}

\address{
Institute of Materials Science, University of Tsukuba,
Tsukuba, Ibaraki 305-8573, Japan
}

\maketitle
\begin{abstract}
The finite-$T$ DMRG method is applied to the one-dimensional
Kondo lattice model to calculate dynamic correlation functions. 
Dynamic spin and charge correlations, 
$S_{\rm f}(\omega)$, $S_{\rm c}(\omega)$, and $N_{\rm c}(\omega)$,
and quasiparticle density of 
states $\rho(\omega)$ 
are calculated in the paramagnetic metallic phase for various 
temperatures and hole densities.
Near half filling, it is shown that
a pseudogap grows in these dynamic correlation functions 
below the crossover temperature characterized by the spin gap 
at half filling. 
A sharp peak at $\omega=0$ evolves at low temperatures in 
$S_{\rm f}(\omega)$ and $N_{\rm c}(\omega)$. 
This may be an evidence of the formation of the collective
excitations, and this confirms that the metallic phase
is a Tomonaga-Luttinger liquid in the low temperature limit.
\end{abstract}

\vskip2pc]

\narrowtext


Heavy fermion systems have attracted much attention for more than
a decade because of their enormous mass renormalization and
diverse ground states including unconventional
superconductivity.\cite{Stew}
The periodic Anderson model (PAM) and the Kondo lattice model (KLM) 
are their canonical models, and
their properties have intensively been investigated.\cite{Hew}
In the framework of the PAM, hybridized band picture with 
strong renormalization is widely accepted for a scenario of the 
formation of heavy quasiparticles.\cite{Col,Ue,Gun,Yama} 
The formation of heavy quasiparticles is  supported by
numerical studies on the PAM in infinite dimensions 
(D=$\infty$)\cite{Jarrell1,Jarrell2}
and in one dimension (1D).\cite{Tsutsui,Eder}
On the other hand it is a nontrivial question 
how the heavy fermion state is formed in the KLM,\cite{ShibaFa}
since the charge degrees of freedom  are completely
suppressed for the localized $f$-electrons. 
For this problem it is important to understand
the relation between the Kondo effect and the hybridization picture.
An important energy scale is the coherence temperature of the
heavy quasiparticles, $T_{\mbox{\scriptsize coh}}$, but several approaches seem to
give controversial results.
Gutzwiller approximation suggests that $T_{\mbox{\scriptsize coh}}$ is enhanced
compared with the single-impurity Kondo temperature 
$T_{\mbox{\scriptsize K}}$,\cite{Ue} 
while Nozi\`eres predicted that $T_{\mbox{\scriptsize coh}}$ is rather suppressed 
due to the exhaustion mechanism.\cite{Noz}
Several numerical calculations on the PAM in D=$\infty$ and 1D 
suggest the suppression of $T_{\mbox{\scriptsize coh}}$.\cite{Jarrell1,Jarrell2,Tsutsui,Eder}

Intersite correlations are believed to play an important role 
for the formation of heavy quasiparticles in real compounds,
but they are not taken into account in D=$\infty$. 
1D systems provide complementary information,
since the intersite correlations generally become more dominant
in lower dimensions.
Precise numerical calculations are also feasible in 1D compared with 2D or 3D.
Therefore the 1D KLM may be another appropriate starting point to obtain reliable 
information on dynamics, although the ground state may not be a Fermi 
liquid.


Ground-state properties of the 1D KLM
have been studied extensively and its magnetic phase
diagram is now determined.\cite{TSU,RMP}
At half filling the ground state is a Kondo spin-liquid insulator 
with a spin gap $\Delta_{\mbox{\scriptsize s}}$,\cite{THUS,YW,NU,Tsve,SNUI,FK}
and the charge gap $\Delta_{\mbox{\scriptsize c}}$ is much
larger than $\Delta_{\mbox{\scriptsize s}}$ due to the
correlation effects on the gap formation.\cite{NU,SNUI}
Upon finite hole doping $\delta$, both spin and charge gaps close and 
the ground state is expected to belong to the universality class
of the Tomonaga-Luttinger liquid (TLL).\cite{UNT,fujimoto,STU}

Thermodynamic properties at finite $\delta$ have been studied 
by the finite-temperature density-matrix renormalization-group
(finite-$T$ DMRG) method\cite{DMRG,Bursill,Wang,Shibata} 
and it is found that there 
exist two crossover temperatures.\cite{ST}
The first crossover is observed as a broad peak in the $T$-dependence 
of the spin susceptibility $\chi_{\mbox{\scriptsize s}}$
when the temperature is lowered from infinity. 
The crossover temperature $T^*_1$ may be defined by its
peak position, and calculations for various $J$'s show that
$T^*_1$ is scaled by $\Delta_{\mbox{\scriptsize s}}$. \cite{SATSU,ST}
The charge susceptibility $\chi_{\mbox{\scriptsize c}}$ and 
$\chi_{\mbox{\scriptsize s}}$ both decrease below $T^*_1$ 
but they turn to increasing again at further low temperatures.
The susceptibilities are expected to finally approach the $T=0$ 
value determined by the velocities of the collective excitations of the 
TLL ground state, and this saturation is actually observed
in $\chi_{\mbox{\scriptsize c}}$ for some $J$ and $\delta$.\cite{ST}
The second crossover temperature $T^*_2$
may be defined to characterize this saturation, and is essentially
determined by either spin or charge velocity depending on the
channel.
The zero temperature susceptibilities calculated 
by the DMRG method show a diverging behavior
as $\delta\rightarrow 0$, and this means vanishing of $T^*_2$.\cite{ST}

In the present work we calculate temperature 
and doping dependence of the dynamic correlation functions and
clarify the character of these crossovers. 
It will be shown that the higher crossover temperature $T^*_1$ 
corresponds to the characteristic temperature of pseudogap
formation in the dynamic spin and charge structure factors.
At the same time a sharp peak structure appears at $\omega=0$, 
indicating the formation of the collective excitations of the TLL at 
low temperatures. The lower crossover temperature $T^*_2$ may 
correspond to the coherence temperature of these collective excitations.

The Hamiltonian of the 1D KLM is described as
\begin{equation}
  {\cal H}= -t \sum_{i,s} \left( c^{\dagger}_{i s}c_{i+1 s} 
       + \mbox{H.c.}\right)
  +J\sum_{i,s,s'}{\bf S}_{i}\cdot{\textstyle\frac{1}{2}}
\mbox{\boldmath $\sigma$}_{ss'} c^{\dagger}_{is}c_{is'}
\label{KLM}
\end{equation}
with standard notations.
The density of conduction electron $n_{\mbox{\scriptsize c}}$ 
is unity at half filling,
and hole doping ($n_{\mbox{\scriptsize c}}=1-\delta$) is 
physically equivalent 
to electron doping ($n_{\mbox{\scriptsize c}}=1+\delta$) 
due to particle-hole symmetry.

\begin{figure}[t]
  \epsfxsize=70mm \epsffile{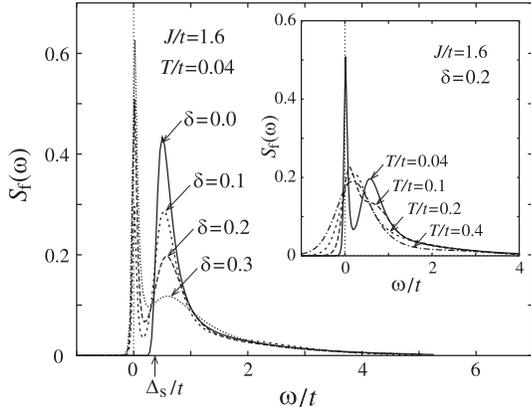}
\caption{
Dynamic spin correlations of the $f$-spins in the 1D KLM.
$\Delta_{\mbox{\scriptsize s}}=0.4t$ is the spin gap at $\delta=0$.
}
\label{Sf}
\end{figure}

\begin{figure}[t]
  \epsfxsize=70mm \epsffile{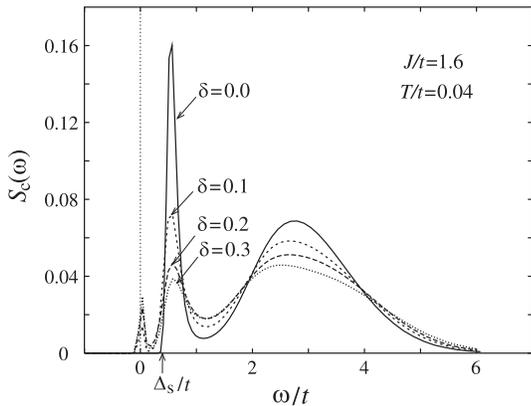}
\caption{
Dynamic spin correlations of the conduction electrons.
}
\label{Sc}
\end{figure}

\begin{figure}
  \epsfxsize=70mm \epsffile{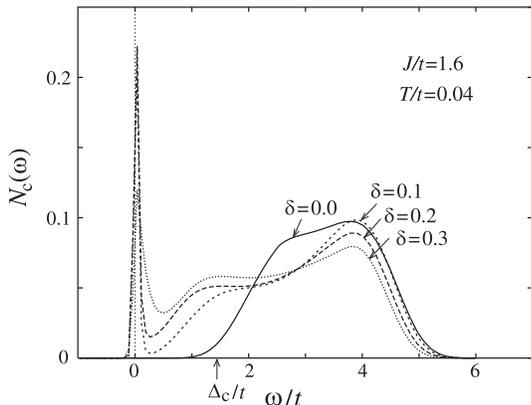}
\caption{
Dynamic charge correlations.
$\Delta_{\mbox{\scriptsize c}}=1.4t$ is the charge gap at $\delta=0$.
}
\label{Nc}
\end{figure}

Dynamic correlation functions of local quantities
can be calculated at finite temperatures for infinite-size systems by
the DMRG method\cite{DMRG} applied to the quantum transfer 
matrix.\cite{Betsu}
Imaginary-time correlation functions are first calculated
from the eigenvector of the maximum eigenvalue of the quantum transfer 
matrix,\cite{Mutou,SU,DW}
and then they are analytically continued to real 
frequency using the maximum entropy method.\cite{Jarrell1}
The advantages of this method are that the finite-$T$ DMRG method 
has no statistical errors and that we 
do not need the extrapolation on the system size.
This approach was first applied to the insulating phase of 
the 1D KLM and the many-body nature of the gap formation 
is revealed.\cite{Mutou,SU}
The present study is the first application to a metallic
phase. In our calculations we usually keep 50 states in the finite-$T$ DMRG
procedure with the Trotter number 60.

We have calculated temperature dependence of several dynamic
correlation functions. 
The results of the local spin dynamics of the $f$-spins,
$S_{\mbox{\scriptsize f}}(\omega)=\int\frac{dq}{2\pi} 
S_{\mbox{\scriptsize f}}(q,\omega)$, 
are shown for $J/t=1.6$ and $\delta=0.2$ in the 
inset of Fig.~\ref{Sf}.
Note that the complete suppression
of charge fluctuations for $f$-spins imposes the sum rule,
$\int \mbox{d}\omega S_{\mbox{\scriptsize f}}(\omega)=1/4$,
independent of $J$, $\delta$ and $T$.
We can see that a peak structure appears around 
$\omega\sim\Delta_{\mbox{\scriptsize s}}=0.4t$ at $T \leq 0.2t$,
and similar peak structure is also observed for different 
Kondo coupling $J/t=1.2$ at $T\leq 0.06t$, where 
$\Delta_{\mbox{\scriptsize s}}=0.16t$.
Based on these results, we may conclude that the
characteristic temperature of the peak formation is scaled by
the higher crossover temperature $T^*_1$ determined by 
$\chi_{\mbox{\scriptsize s}}(T)$.
At the same time another peak structure grows at $\omega=0$, 
when $\delta$ is finite.  It suggests the formation of 
the collective spin excitations of the TLL at low temperatures.
We expect that the low energy part finally approaches 
$|\omega|^{K_c}$, which is 
predicted by the TLL theory at $T=0$.\cite{HS}

The doping dependence of $S_{\mbox{\scriptsize f}}(\omega)$ is 
shown at the low temperature $T=0.04t \ll \Delta_{\mbox{\scriptsize s}}$
in the main panel of Fig.~\ref{Sf}.
With increasing $\delta$, the peak intensity at $\omega=0$ grows,
while the intensity around $\omega=\Delta_{\mbox{\scriptsize s}}$ 
is reduced as a consequence of the sum rule discussed before.
We have checked that the peak intensity at $\omega=0$ is $0.25 \delta$
within a few percent. 
The energy scale estimated from the peak width at $\omega=0$
at $T=0.04t$ is smaller than the lowest temperature in our calculations.
This means the weight around $\omega=0$ yields nearly free spin degrees of 
freedom with density $0.25 \delta$ down to around this temperature $T\sim 0.04t$.
This is consistent with the behavior of the static spin susceptibility,
$\chi_{\mbox{\scriptsize s}}=\delta/(4T)$, observed at low temperatures.\cite{ST}

In contrast to $S_{\mbox{\scriptsize f}}(\omega)$ 
the $\omega=0$ peak is small in the conduction-electron
spin correlations, $S_{\mbox{\scriptsize c}}(\omega)=
\int\frac{dq}{2\pi}S_{\mbox{\scriptsize c}}(q,\omega)$, 
as shown in Fig.~\ref{Sc}. 
This shows that the low-energy part of the spin 
degrees of freedom of the conduction electrons are nearly exhausted  
to form spin singlet with $f$-spins and
thus a clear peak is formed at $\omega=\Delta_{\mbox{\scriptsize s}}$.
Therefore, the low energy spin dynamics is mostly dominated
by the $f$-spin degrees of freedom.
However, the intensity of the peak around 
$\omega\sim \Delta_{\mbox{\scriptsize s}}$
is less than $1/3$ of the corresponding peak in 
$S_{\mbox{\scriptsize f}}(\omega)$ and over a half of the 
total weight extends over higher frequencies $t\le \omega\le 5t$.
This means that although the low energy part is dynamically coupled 
with $f$-spins, the majority part of the conduction spin degrees of freedom
have another energy scale of almost the band width $\sim 4t \gg \Delta_{\mbox{\scriptsize s}}$ 
when the Kondo coupling is small $J<4t$.
This shows that only the conduction electrons close to the Fermi level
screen the $f$-spins as pointed out by Nozi\`eres.\cite{Noz}
Rather surprisingly, 
even though the screening by conduction electrons is not complete,
the intensity at $\omega=0$ in $S_{\mbox{\scriptsize f}}(\omega)$ is  
almost $\delta/4$, just like the $J=\infty$ case, where each
conduction electron screens one $f$-spin.
This shows the importance of the $f$-$f$ spin correlations
for the formation of the Kondo singlet state.

The doping dependence of the dynamic charge correlations, 
$N_{\mbox{\scriptsize c}}(\omega)=\int\frac{dq}{2\pi}
 N_{\mbox{\scriptsize c}}(q,\omega)$, is shown in Fig.~\ref{Nc}. 
At half filling, the charge excitations 
are exponentially suppressed below the crossover temperature $T^*_1$
at $\omega<\Delta_{\mbox{\scriptsize c}}$.\cite{SATSU}
Upon hole doping, a sharp peak appears at $\omega=0$ as in 
$S_{\mbox{\scriptsize f}}(\omega)$. This indicates
the formation of collective charge excitations of the TLL
as discussed for $S_{\mbox{\scriptsize f}}(\omega)$.
The peak intensity increases with $\delta$,
and this means that the effective carrier density of the TLL is 
strongly renormalized from $n_{\mbox{\scriptsize c}}=1-\delta$.
The renormalization of carrier is naturally explained in the limit of 
strong $J$, where each conduction electron forms a local singlet with
the $f$-spin on the same site.
In this limit, effective carriers are introduced by hole
doping, and their main component is the unscreened $f$-spins with 
density $\delta$,\cite{UNT} as discussed for $S_{\mbox{\scriptsize f}}(\omega)$.
The TLL theory predicts that the low-energy asymptotic form of 
$N_{\mbox{\scriptsize c}}(\omega)$ is 
$|\omega|^{\mbox{\scriptsize min}(K_c,4K_c-1)}$
near $\omega=0$, and we expect that the $\omega=0$ peak finally 
approaches this form in the low temperature limit.

\begin{figure}
  \epsfxsize=70mm \epsffile{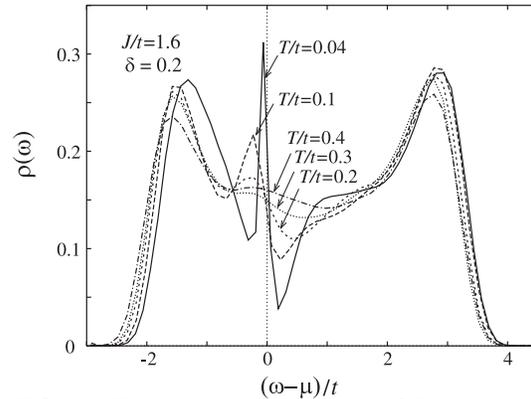}
\caption{
Temperature dependence of the quasiparticle density of states.
}
\label{DOS1}
\end{figure}

We now consider crossover behavior of the quasiparticle
density of states, $\rho(\omega)$. Figure \ref{DOS1} shows the
temperature dependence at $\delta=0.2$ and $J/t=1.6$.
We can see that a pseudogap develops
just above the Fermi level $\omega=\mu$ below $T/t \sim 0.4$.
Similar behavior is observed also for $J/t=1.2$ below $T/t \sim 0.15$.
Based on these results,
we may conclude that the characteristic temperature of pseudogap 
formation is scaled by $T^*_1$ defined from 
$\chi_{\mbox{\scriptsize s}}(T)$.
This conclusion is consistent with the results obtained 
at half filling.\cite{Mutou,SU}

The crossover behavior around $T^*_1$ may be explained as follows. 
Below $T^*_1$ thermal fluctuations of the $f$-spins are substantially 
suppressed, since the temperature is lower than the characteristic 
energy scale of the $f$-spin excitations $\Delta_{\mbox{\scriptsize s}}$.
When $\delta$ is small, the characteristic time scale of the
dominant part of the $f$-spins is given by $\Delta_{\mbox{\scriptsize s}}^{-1}$, 
and this is much longer than the time scale of quasiparticle propagation
$\tau_{\mbox{\scriptsize qp}}$, since $\tau_{\mbox{\scriptsize qp}}$ 
may be determined by the inverse of
bare hopping energy and charge gap, $t^{-1}$ and 
$\Delta_{\mbox{\scriptsize c}}^{-1}$.
Therefore, concerning the quasiparticle excitations, the $f$-$f$
spin correlations may be assumed to be static.  Because of
their staggered nature in space, these almost static $f$-$f$ spin
correlations induce the opening of a gap in $\rho(\omega)$.
Of course, as shown in $S(\omega)$ and $N(\omega)$, both the 
$f$-spins and the conduction electrons 
have a slow dynamics when $\delta$ is finite, and therefore the 
gap discussed here is not a real gap but
rather a pseudogap for finite $\delta$.
Thus the pseudogap develops below $T^*_1$.

\begin{figure}
  \epsfxsize=70mm \epsffile{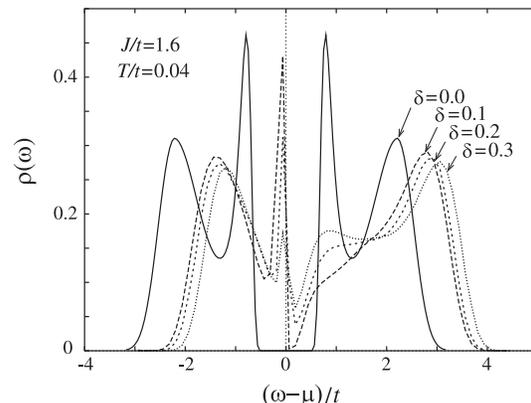}
\caption{
Doping dependence of the quasiparticle density of states.
Quasiparticle gap at $\delta=0$ is $\Delta_{\mbox{\scriptsize qp}}=0.7t$.}
\label{DOS2}
\end{figure}

We next consider doping dependence to study small structures in
$\rho(\omega)$ near the Fermi level $\omega=\mu$.
The results for $J/t=1.6$ at $T/t=0.04$ are shown in Fig.~\ref{DOS2}
at various dopings.
The pseudogap becomes more prominent with approaching
half filling, where the clear quasiparticle gap, 
$\Delta_{\mbox{\scriptsize qp}}/t=0.7$, exists. 
The sharp peak structure just below $\omega=\mu$ also grows 
with decreasing $\delta$ and seems to continuously connect to the 
gap edge structure at $\delta=0$.

The nature of the structure near the Fermi level is not fully
understood yet.  One possible scenario is the mean-field type
argument assuming the $f$-spin helical SDW order with wave number
$2k_F=\pi(1-\delta)$.  
Band mixing induces gap opening at $k=\pi (\pm 1+\delta)/2$,
and two new van-Hove (vH) divergent singularities appear in
each of the two hybridized bands.  The Fermi level sits
between these two new singularities in the lower hybridized
band, as far as $\delta$ is small.  A slightly different
scenario is that the gap is induced by the short-range
$f$-spin correlations with wave number $\pi$.  Then, there
appears only one vH singularity in each band.  The peak-like
behavior observed in Fig.~\ref{DOS2} may be identified as the lower
vH singularity in
\begin{figure}[t]
  \epsfxsize=70mm \epsffile{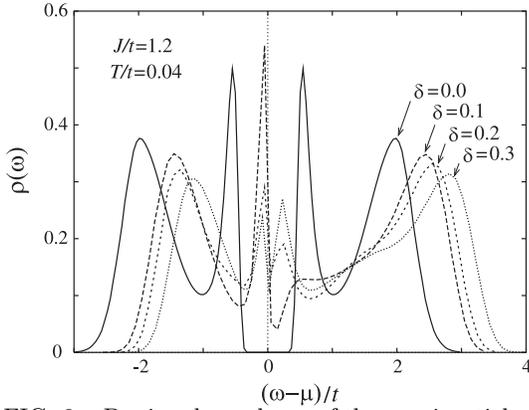}
\caption{
Doping dependence of the quasiparticle density of states.
Quasiparticle gap at $\delta=0$ is $\Delta_{\mbox{\scriptsize qp}}=0.47t$.}
\label{DOS3}
\end{figure}
\noindent
the first scenario, while in the second
one it is a consequence of the shift of the Fermi level
towards the top of the lower band as $\delta \rightarrow 0$.
However, it is not straightforward to describe
the behavior of $\rho(\omega)$ just above the Fermi level
in terms of these pictures.
Another quite different scenario is that this peak is due
to the Kondo singlet formation and reminiscence of the
Kondo resonance.  This may also be expressed as the renormalized
hybridized bands as proposed for the PAM, with the Fermi level
inside the lower band.  The
renormalized hybridized bands are actually observed for the PAM
in D=$\infty$\cite{Dinf} and 1D.\cite{Eder}  Although
the results of the 1D PAM indicate a quite symmetric
$\rho(\omega)$ near $\omega=\mu$ for conduction electrons, which
differs from our results, there is a considerable asymmetry
in the results for the D=$\infty$ PAM, similar to our results.
With decreasing $\delta$, the Fermi level shifts towards the top
of the lower band, and $\rho(\omega)$ near the Fermi level grows
accordingly, which is consistent with the behavior in Fig.~\ref{DOS2}.
Therefore, this scenario is quite promising.
Anyway, the slow dynamics of $f$-spin correlations may be important to 
understand the low energy dynamics of quasiparticle motion, 
and we will make further investigations in future study.

As the temperature decreases down to zero, the 
interaction between the effective carriers 
becomes relevant and the low energy excitations are
expected to be described as a TLL.\cite{fujimoto,STU} 
The second crossover temperature is defined to characterize this
behavior, as discussed before.
Figure \ref{DOS3} shows $\rho(\omega)$ for the smaller coupling $J/t=1.2$.
In this case, the second crossover temperature is 
relatively high compared with the case of $J/t=1.6$,\cite{ST}
and the asymptotic TLL behavior may be observed.
For $\delta=0.2$ and $0.3$, a dip structure is now observed at the 
Fermi level $\omega=\mu$, which is absent in Fig.~\ref{DOS2}. 
It is known that the TLL has $\rho(\omega)$ with 
a dip structure at the Fermi level
as $\rho(\omega)\sim |\omega-\mu|^\alpha$ at $T=0$ where 
$\alpha=(K_{\mbox{\scriptsize c}}-1)^2/(4K_{\mbox{\scriptsize c}})$.\cite{HS} 
Here $K_{\mbox{\scriptsize c}}$ is the Luttinger liquid parameter and less than $1/2$
in the 1D KLM.\cite{STU} 
Thus the dip structure in Fig.~\ref{DOS3} is consistent with
the TLL picture.
Such dip structure is not found for small doping $\delta=0.1$. 
This agrees with the previous study of the thermodynamics in the 
1D KLM,\cite{ST}
which shows that $T^*_2$ is lower for smaller $\delta$
and vanishes as $\delta \rightarrow 0$.

To summarize we have calculated dynamic quantities at various 
temperatures and hole densities,
and clarified characters of the two crossovers in 
the paramagnetic metallic phase. Below the first crossover temperature 
$T^*_1$ it has been shown that a
pseudogap develops in the density of states and dynamic 
correlation functions, and 
$S_{\mbox{\scriptsize f}}(\omega)$ and $S_{\mbox{\scriptsize c}}(\omega)$
both show a peak
structure around $\omega=\Delta_{\mbox{\scriptsize s}}$ 
as in the half-filling case. At the same time a peak structure 
appears at $\omega=0$ in $S_{\mbox{\scriptsize f}}(\omega)$ 
and $N_{\mbox{\scriptsize c}}(\omega)$,
and its intensity increases with hole density $\delta$.
The $\omega=0$ peak indicates that as a consequence of local
Kondo singlet formation effective carriers are strongly
renormalized to have density $\delta$ and small energy scale.
The increase of the peak intensity with $\delta$ is 
naturally explained in the limit of strong $J$,
where effective carriers of the TLL are the unscreened $f$-spins
whose density is $\delta$. We note that this
is consistent with the large Fermi surface.\cite{UNT,fujimoto,STU}
Below the second crossover temperature $T^*_2$, the interaction between 
the effective carriers becomes relevant and the renormalized carriers 
are expected to evolve into a TLL.  This is supported by several sets of the
present results.


\end{document}